\documentclass[12pt]{iopart}

\usepackage{iopams}  
\usepackage{graphicx}
\usepackage{cite}

\begin{document}

\title{
Ultrasonic observation of small Fermi surfaces in La$T$In$_5$ ($T$ = Co, Rh, Ir)
}

\author{
Ryosuke Kurihara$^{1*}$,
Yusuke Hirose$^2$,
Kazuki Matsui$^3$,
Atsushi Miyake$^4$,
Ryoma Tsunoda$^5$,
Masaki Kondo$^3$,
Rikio Settai$^6$,
Mitsuhiro Akatsu$^6$,
Yuichi Nemoto$^5$,
Hiroshi Yaguchi$^1$,
and
Masashi Tokunaga$^3$
}

\address{
$^1$Department of Physics and Astronomy, Tokyo University of Science, Noda, Chiba 278-8510, Japan
\\
$^2$Materials Sciences Research Center, Japan Atomic Energy Agency, Tokai, Ibaraki 319-1195, Japan.
\\
$^3$The Institute for Solid State Physics, The University of Tokyo, Kashiwa, Chiba 277-8581, Japan
\\
$^4$Institute for Material Research, Tohoku University, Oarai, Ibaraki 311-1313, Japan
\\
$^5$
Graduate School of Science and Technology, Niigata University, Ikarashi, Niigata 950-2181, Japan
\\
$^6$
Faculty of Science, Niigata University, Ikarashi, Niigata 950-2181, Japan
}

\vspace{10pt}
\begin{indented}
\item[]$^*$ E-mail : r.kurihara@rs.tus.ac.jp or iron.pnictide.man@gmail.com
\end{indented}

\begin{abstract}
We performed high-field ultrasonic measurements on La$T$In$_5$ ($T$ = Co, Rh, Ir) to reveal the origin of the small Fermi surface that was recently observed in LaRhIn$_5$ with an oscillation frequency of 6.8 T.
We observed quantum oscillations originating from this Fermi surface in LaRhIn$_5$.
In addition, we revealed that LaCoIn$_5$ and LaIrIn$_5$ exhibit quantum oscillations with frequencies below 100 T, indicating hidden Fermi surfaces in these compounds. 
Furthermore, Co-substituted LaRhIn$_5$ exhibited quantum oscillations with a frequency of 10 T.
Our results suggest that the small Fermi surface originates from bulk properties and that $3d$ electrons of the transition metal contribute to its formation.
\end{abstract}

%
%
%
%
%

\section{
\label{Introduction}
Introduction
}

In quantum limit states realized under high magnetic fields, novel phase transitions originating from electron correlations have been proposed, including excitonic transitions 
\cite{Halperin_JJAP26}.
In addition, the realization of an excitonic-insulator transition accompanying structural phase transitions due to electron-electron and electron-phonon interactions has been discussed
\cite{Sugimoto_PRB93}.
Therefore, small Fermi surfaces which are caused by Dirac-like band dispersion can provide the crystal symmetry breaking above quantum limit states because strong electron-phonon interactions have been proposed in such systems
\cite{Schindler_PRB102}.

LaRhIn$_5$ is one of the good materials for this purpose.
Several de Haas-van Alphen (dHvA) and Shubnikov-de Haas (SdH) effects studies have suggested that a small Fermi surface with a frequency of 7 T
\cite{Goodrich_PRL89, Guo_NatCommun12},
as well as other Fermi surfaces with frequencies over 1 kT
\cite{Shishido_JPSJ71},
contribute to its physical properties under magnetic fields.
The small Fermi surface with a frequency of 7 T is considered to be related to the linear band dispersion and a nontrivial Berry phase
\cite{Mikitik_PRL93, Guo_NatCommun12}.
Band calculations have reproduced the large Fermi surface of LaRhIn$_5$
\cite{Shishido_JPSJ71}.
However, the calculated shape of the small Fermi surface is not consistent with experimental results.

To deepen our understanding of the small Fermi surface, we focused on the high-field ultrasonic measurements in LaRhIn$_5$ and related materials, such as La$T$In$_5$ ($T =$ Co, Rh, Ir).
Ultrasonic measurements are sensitive techniques for detecting quantum oscillations in the elastic constant of materials
\cite{Mase_JPSJ31, Suzuki_JMMM63, Settai_JPSJ63}.
Furthermore, this technique can be used for high-field quantum oscillation measurements with a pulse magnet
\cite{Akiba_PRB98, Kurihara_PRB101, Kondo_PRB107}.
LaRhIn$_5$ and its related compounds exhibit high electrical conductivity, therefore, to measure the SdH effect with high precision, it is necessary to fabricate microstructures and to increase the resistance value
\cite{Guo_NatCommun12}.
On the other hand, we can measure quantum oscillations using ultrasonic measurements with large-size single crystals.
Understanding the transition-metal dependence of the quantum oscillations in La$T$In$_5$ can further clarify the origin of the small Fermi surface.

This paper is organized as follows.
In Sec. \ref{sect_exp}, the sample preparation and experimental procedures are described.
In Sec. \ref{Result}, we present that quantum oscillations with a frequency below 100 T are observed in the magnetic field dependence of the elastic constant in La$T$In$_5$ compounds.
In Sec. \ref{Discussion}, we discuss the material dependence of the observed small Fermi surfaces.
We present the conclusion based on our results in Sec. \ref{conclusion}.

\section{
\label{sect_exp}
Experimental methods
}

Single crystals of La$T$In$_5$ ($T$ = Co, Rh, Ir) were grown by the flux method
\cite{Shishido_JPSJ71}.
Crystallographic orientations were determined using the Laue backscattering method.

The ultrasonic pulse-echo method with a numerical vector-type phase detection technique was used for the ultrasonic velocity $v_{ij}$ and the phase difference $\phi_{ij} = 2 \pi \left( 2n-1 \right) l f /v_{ij}$ where $n$ is the index of $n$th ultrasonic echo signal, $l$ is the sample length, and $f$ is the ultrasonic frequency 
\cite{Luthi Phys. Ac.}.
Piezoelectric transducers using LiNbO$_3$ plates with a 36$^\circ$ Y-cut and an X-cut (YAMAJU CO) were employed to generate longitudinal ultrasonic waves with the fundamental frequency of approximately $f$ = 30 MHz and the transverse waves with 18 MHz, respectively.
The elastic constant, $C_{ij} = \rho v_{ij}^2$, was calculated from the ultrasonic velocity, $v_{ij}$, and the mass density, $\rho = 8.17$ g/cm$^3$ of LaRhIn$_5$
\cite{Hieu_JPSJ76}
and $\rho = 8.84$ g/cm$^3$ of LaIrIn$_5$
\cite{Macaluso_JSSC166}.
On the other hand, we could not determine the absolute value of the elastic constant of LaCoIn$_5$ and Co-substituted LaRhIn$_5$ because the sample size was not enough to estimate the sound velocity.
Thus, the field dependence of the phase difference, denoted as $\mathit{\Delta}\phi_{ij} = \phi_{ij}\left( H = 0 \right) -\phi_{ij} \left( H \right)$, is shown.
The ultrasonic propagation direction, $\boldsymbol{q}$, and the polarization direction, $\boldsymbol{\xi}$, for the elastic constant $C_{ij}$ and the phase difference $\phi_{ij}$ and the magnetic field direction, $\boldsymbol{H}$, are shown in the figures in the present paper.

The magnetic field dependence of the elastic constant in LaRhIn$_5$ was measured by Physical Property Measurement System (Quantum Design, PPMS) at the International MegaGauss Science Laboratory, ISSP.
For measurements in magnetic fields of up to 55 T on LaCoIn$_5$, LaIrIn$_5$, and Co-substituted LaRhIn$_5$, a non-destructive pulse magnet with a time duration of 36 ms installed at The Institute for Solid State Physics, The University of Tokyo, was used
\cite{Miyata_IEEE36}.
$^4$He cryostat was used to obtain low temperatures down to 1.4 K.


\section{
\label{Result}
Results
}

\begin{figure}[b]
\centering
\includegraphics[clip,width=0.9\textwidth]{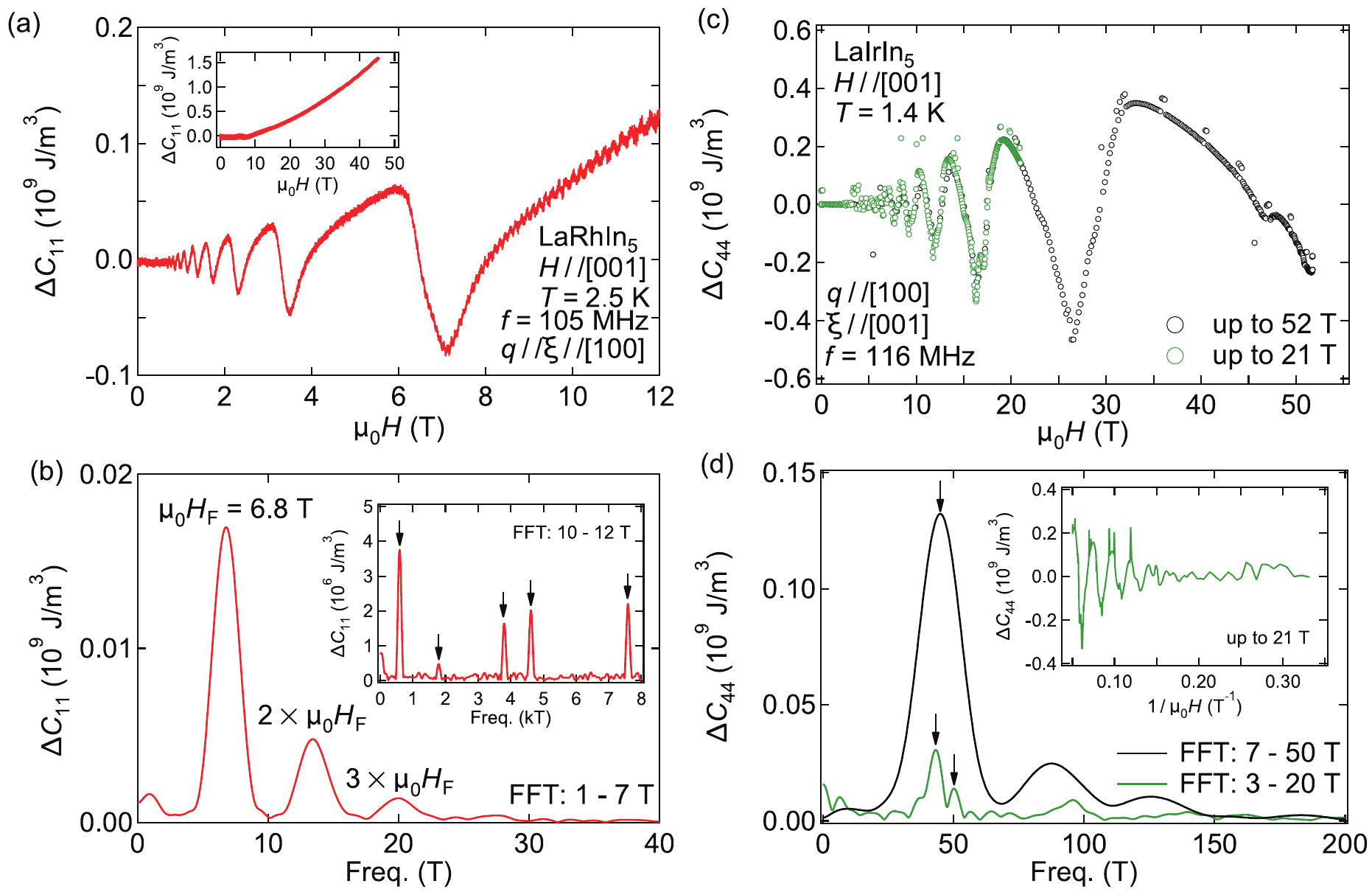}
\caption{
(a) Magnetic field dependence of the longitudinal elastic constant $\mathit{\Delta}C_{11} = C_{11}\left( H \right) - C_{11}\left( H = 0 \right)$ at 2.5 K for $\boldsymbol{H}//[001]$.
(b) Power spectra of the fast Fourier transform of $\mathit{\Delta}C_{11}$ in the range of 1 to 7 T.
The inset in panel (b) shows the FFT spectra in the range of 10 to 12 T.
The vertical arrows indicate oscillation frequencies of 0.6, 1.8, 3.7, 4.6, and 7.6 kT.
(c) Magnetic field dependence of the transverse elastic constant $\mathit{\Delta}C_{44} = C_{44}\left( H \right) - C_{44}\left( H = 0 \right)$ at 1.4 K for $\boldsymbol{H}//[001]$.
$\mathit{\Delta} C_{44}$ indicated by the black (green) open circles is measured by the pulsed magnetic fields with a maximum field of 52 T (21 T).
(d) Power spectra of the fast Fourier transform of $\mathit{\Delta}C_{44}$.
The inset in panel (d) shows the inverse magnetic field dependence of $\mathit{\Delta} C_{44}$ measured by the pulsed magnetic fields with a maximum field of 21 T in the range of 3 to 20 T.
The vertical arrows indicate oscillation frequencies of 45.0 T (43.2 and 50.2 T) for the FFT range of 7 to 50 T(3 to 20 T).
}
\label{Fig1}
\end{figure}

Figure \ref{Fig1}(a) shows the magnetic field dependence of the longitudinal elastic constant $C_{11}$ of LaRhIn$_5$ for $\boldsymbol{H}//[001]$ at 2.5 K.
We observed quantum oscillations in $C_{11}$.
The oscillation frequency of $\mu_0 H_\mathrm{F} = 6.8$ T is determined by the FFT spectra of $\mathit{\Delta} C_{11} = C_{11} \left(H \right) - C_{11} \left(H = 0 \right)$ [see Fig. \ref{Fig1}(b)].
This value of $\mu_0 H_\mathrm{F}$ is consistent with the oscillation frequency determined by dHvA and SdH measurements
\cite{Goodrich_PRL89, Guo_NatCommun12}.
In addition to at $\mu_0 H_\mathrm{F}$, the FFT spectra of $\mathit{\Delta} C_{11}$ show several peaks at $2 \times \mu_0 H_\mathrm{F}$ and $3 \times \mu_0 H_\mathrm{F}$, which is consistent with the higher harmonics of the fundamental value of $\mu_0 H_\mathrm{F} = 6.8$ T.
We also observed quantum oscillations with frequencies over 0.5 kT [see the inset in Fig. \ref{Fig1}(b)]. 
These frequencies are consistent with the previous studies
\cite{Shishido_JPSJ71}.

To investigate the field-induced crystal symmetry breaking in LaRhIn$_5$ in the quantum limit, we measured $C_{11}$ in fields of up to 45 T [see the inset in Fig. \ref{Fig1}(a)].
$\mathit{\Delta} C_{11}$ shows the monotonic increase in the field above 7 T, indicating that the crystal symmetry breaking related to the longitudinal elastic constant $C_{11}$ is absent up to 45 T. 
Comprehensive measurements of the other elastic constants and further high-field measurements can be necessary.

Figure \ref{Fig1}(c) shows the magnetic field dependence of the transverse elastic constant $C_{44}$ of LaIrIn$_5$ for $\boldsymbol{H}//[001]$ at 1.4 K up to 52 T.
$C_{44}$ up to 52 T exhibits quantum oscillations with a frequency of 45.0 T [see Fig. \ref {Fig1}(d)].
In addition to the peak at 45.0 T, the FFT spectra show peaks around 90 and 130 T.
Since a sawtooth shape appears in the quantum oscillations in $C_{44}$ and such a shape is observed for LaRhIn$_5$, we deduce that the peaks at 90 and 130 T correspond to the higher harmonics of the fundamental frequency of 45.0 T.

In addition to the sawtooth shape, the field dependence of $C_{44}$ exhibits other oscillation-like anomalies around 22 and 45 T [see Fig. \ref{Fig1}(c)]. 
To investigate the of existence of other small Fermi surfaces, we performed the field dependence of $C_{44}$ up to 21 T [see black (green) open circles in Fig. \ref{Fig1}(c)].
As shown in Fig. \ref{Fig1}(d), the calculated FFT spectra of the quantum oscillations in $\mathit{\Delta} C_{44}$ show two peaks at 43.2 and 50.2 T.
This result indicates that LaIrIn$_5$ has two small Fermi surfaces.
The two-frequency contribution to the quantum oscillations in $C_{44}$ is also indicated by the beating behavior shown in the inset of Fig. \ref{Fig1}(d).

\begin{figure}[t]
\centering
\includegraphics[clip,width=0.9\textwidth]{Fig2.pdf}
\caption{
(a) Magnetic field dependence of the phase difference $-\mathit{\Delta}\phi_\mathrm{L}$ of the in-plane longitudinal elastic constant of LaCoIn$_5$ at several temperatures for $\boldsymbol{H}//[001]$.
The data sets are shifted consequently along the $-\mathit{\Delta}\phi_\mathrm{L}$ axes for clarity.
(b) Magnetic field dependence of the phase difference $-\mathit{\Delta}\phi_\mathrm{L[110]}$ of the in-plane longitudinal elastic constant of LaCo$_{0.1}$Rh$_{0.9}$In$_5$  at 1.4 K for $\boldsymbol{H}//[001]$.
The inset in panel (b) shows the inverse magnetic field dependence of $-\mathit{\Delta}\phi_\mathrm{L[110]}$.
}
\label{Fig2}
\end{figure}

Figure \ref{Fig2}(a) shows the magnetic field dependence of the phase difference $-\mathit{\Delta}\phi_\mathrm{L} = \phi_\mathrm{L}\left( H = 0 \right) -\phi_\mathrm{L} \left( H \right)$ for the in-plane longitudinal elastic constant of LaCoIn$_5$ for $\boldsymbol{H}//[001]$ at several temperatures.
$\mathit{\Delta}\phi_\mathrm{L}$ at 1.4 K shows quantum oscillations.
As the temperature is increased, the amplitude of quantum oscillations seems to be suppressed.
Because of the experimental resolution, we could not calculate the fast Fourier transformation amplitude of the quantum oscillations, but estimated an oscillation frequency of about 35 T from the period of peaks and dips.
For detailed investigation of the quantum oscillations, further experiments are necessary. 

Figure \ref{Fig2}(b) shows the magnetic-field dependence of the phase difference $-\mathit{\Delta}\phi_\mathrm{L[110]} = \phi_\mathrm{L[110]}\left( H = 0 \right) -\phi_\mathrm{L[110]} \left( H \right)$ for the in-plane longitudinal elastic constant of LaCo$_{0.1}$Rh$_{0.9}$In$_5$ for $\boldsymbol{H}//[001]$ at 1.4 K.
We observed the quantum oscillations with a frequency of approximately 10 T [see the inset in Fig. \ref{Fig2}(b)].

\section{
\label{Discussion}
Discussion
} 

\begin{figure}[t]
\centering
\includegraphics[clip,width=0.6\textwidth]{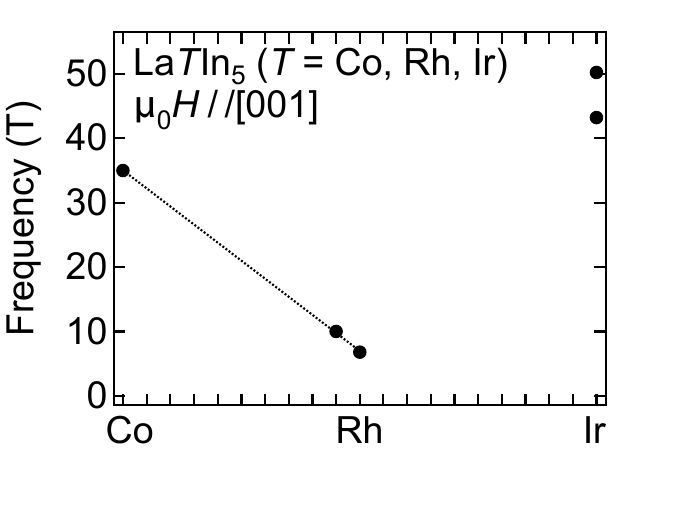}
\caption{
Material dependence of the oscillation frequency of the small Fermi surface in La$T$In$_5$ ($T = $ Co, Rh, Ir) for $\boldsymbol{H}//[001]$.
The dotted line indicates the linear fit of the oscillation frequencies of LaCo$_{1-x}$Rh$_x$In$_5$.
}
\label{Fig3}
\end{figure}

Our ultrasonic experiments have shown that, in addition to LaRhIn$_5$, noth both of LaCoIn$_5$ and LaIrIn$_5$ have Fermi surfaces smaller than 100 T.
These frequencies $\mu_0 H_\mathrm{F}$ are summarized in Fig. \ref{Fig3}.
We deduce that these Fermi surfaces have not been observed in previous dHvA measurements 
\cite{Hall_PRB79, Forzani_DrThesis}.

Based on these results, we discuss the origin of the small Fermi surfaces.
The elastic constants measured by the ultrasound are thermodynamic quantities.
Therefore, the observed quantum oscillations that reflect the properties of the small Fermi surface in La$T$In$_5$ are attributed to its bulk and intrinsic properties. 
In other words, contributions from surface bands and impurities to the quantum oscillations can be ruled out.

The transition metal dependence of the quantum oscillations also plays a key role in understanding the small Fermi surface.
Due to Co substitution for Rh in LaRhIn$_5$, the Fermi levels characterized by the oscillation frequencies $\mu_0 H_\mathrm{F}$ can change from 6.8 T to 10 T.
Taking into account $\mu_0 H_\mathrm{F} \sim 35$ T of LaCoIn$_5$, we can describe $\mu_0 H_\mathrm{F}$ of LaCo$_x$Rh$_{1-x}$In$_5$ by the linear relationship $\mu_0 H_\mathrm{F} = 35.0 - 28.0x$ T.
This result suggests that Co-3$d$ and Rh-$4d$ electrons contribute to the formation of the small Fermi surfaces.
To confirm these assumptions, the detailed Co-substitution dependence of the quantum oscillations, estimating the cyclotron mass, the Berry phase contribution to the quantum oscillations, and band calculations for the small Fermi surfaces in LaCo$_x$Rh$_{1-x}$In$_5$ compounds are necessary.
Investigating quantum oscillations in Ir-substituted LaRhIn$_5$ and Co-substituted LaIrIn$_5$ can confirm Ir-$5d$ contributions to the small Fermi surfaces.

\section{
\label{conclusion}
Conclusion
}

Using bulk-sensitive ultrasonic measurements, we observed quantum oscillations with frequencies of 43.2 and 50.2 T in LaIrIn$_5$ and approximately 35 T in LaCoIn$_5$ in addition to that of 6.8 T in LaRhIn$_5$.
Co-substituted LaRhIn$_5$ also exhibited quantum oscillations with a frequency of approximately 10 T, indicating that Co-$3d$ and Rh-$4d$ electrons are involved in forming the small Fermi surfaces.
Our results can be crucial for understanding the small Fermi surface in LaRhIn$_5$ and its characteristic properties, including the contributions of the Berry phase and the linear band dispersion to the quantum oscillations.

\section*{
Acknowledgments
}

This work was supported by supported by JSPS transformative research areas (A), section (II) (JP 23H04862, 24H01629) and early-career scientists (JP 20K14404).
This work was also partly supported by JSPJ early-career scientists (JP 20K14414).



\end{document}